# Increased Secret Key Throughput in Twin Field Quantum Key Distribution using 4×4 Beam Splitter Detection Network

Ishan Pandey* and Varun Raghunathan
*Department of Electrical Communications Engineering*
*Indian Institute of Science*
*Indian Space Research Organization**
*Bengaluru, India*
ishanp@iisc.ac.in, varunr@iisc.ac.in

*Abstract*—Twin Field Quantum key Distribution (TF-QKD) has attracted recent interest due to the higher secret key capacity better than the fundamental repeaterless limit and extending the achievable distance. The key generation in TF-QKD is based on the post selection of randomized phase slices. This paper describes a technique for enhancing the probability of choosing the phase slices by using four detectors at Charlie's end placed after a 4x4 port beam-splitter network. Using theoretical modelling of secret key-rate and simulations using Strawberry-Fields, we observe an increase in secret key throughput when compared to conventional TF-QKD.

*Keywords—Quantum Communication, Quantum Key Distribution, Quantum communication protocols*

## I. INTRODUCTION

Present day cryptography solutions rely extensively on computationally complex problems which are difficult to break in a reasonable amount of time. Advancements in computational hardware and algorithms, especially a quantum computers pose serious threat to these computation complexity assumptions [1]. Quantum key distribution (QKD) protocols aim to leverage the inherent laws of quantum mechanics rendering them unconditionally secure when implemented with ideal devices obeying the underlying principles [1]. Nevertheless, implementing QKD protocols with single-photon sources and ideal detectors with 100% efficiency and zero dark count is challenging in practice. Additionally, channel losses cannot be assumed to be negligible, especially when operating at metro-haul and long-haul distances. These non-idealities constrain the secret key rate and achievable transmission distance. A fundamental upper bound has been derived for the achievable secret key rate ($R_{max}$) without the use of a repeater, known as the Pirandola-Laurenza-Ottaviani-Banchi (PLOB) Bound [2], given by: $R_{max} = \log_2(1 − \eta_{trans})$ where $\eta_{trans}$ represents the channel transmittance.

The use of coherent sources in QKD protocols introduces vulnerability due to photon number splitting (PNS) attacks, which is often mitigated through the implementation of decoy states [3]. Other potential detector-side, side-channel attacks stemming from non-idealities at the measurement end are addressed through Measurement Device Independent (MDI) QKD protocol [4]. The MDI-QKD protocol makes the measurement party privy to the measurement outcomes without revealing the actual transmitted qubits, which Alice and Bob can decipher based on the announced results. The risk of phase remapping attacks are also mitigated through phase randomization [5]. Despite adhering to the constraints of the PLOB bound, the secret key rate of these protocols typically remains below this limit due to the requirement of coincidence measurement at the middle party.

Twin Field Quantum Key Distribution (TF-QKD) protocol [6] surpasses the PLOB bound limit, encompassing the feature of side channel attack proof of MDI-QKD with additional benefit of extending transmission distances by leveraging single-photon interference. This paper proposes enhancing the secret-key throughput in TF-QKD by replacing the conventional single beam-splitter and two-detector setup with a 4x4 port beam-splitter network and four detectors. This modification increases the secret key rate by ~2.3 times compared to the conventional TF-QKD setup.

This paper is organized as follows: Section II covers the conventional Twin Field Quantum Key Distribution (TF-QKD) protocol, including the secret key rate and the Binary Symmetric Erasure Channel model. Section III details the proposed 4-detector setup, its use in TF-QKD, and its enhanced capability to detect beam-splitting attacks. Section IV describes the simulation setup with the Strawberry Fields package and compares the results with the theoretical expression of the sifted key.

## II. TF-QKD protocol

### A. Protocol Steps

A schematic of the convention TF-QKD protocol is illustrated in Fig. 1 and the essential features of the encoding and decoding scheme described below:

*1) Initialization:* Three nodes are involved in the key distribution, namely Alice, Bob who wish to communication,

The authors acknowledge financial support from Ministry of Electronics and Information Technology (MEITY), Center of Excellence in Quantum Technologies at IISc and DRDO Industry Academia Centre of Excellence, IIT Delhi (DIA-CoE, IITD).

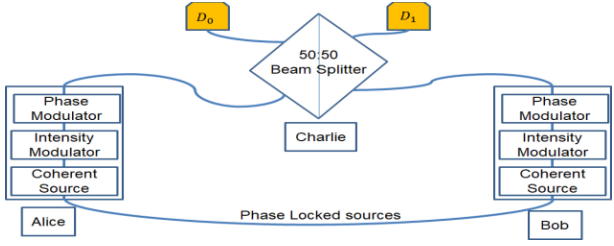

Fig. 1. Schematic of Twin Field QKD implementation.

and Charlie who does the measurements and announces the measurement outcomes. Initially, Alice and Bob agree to divide the phase interval (0 to 2π) into M equal slices $\Delta_k = 2\pi k/M$, where k={0, 1, 2..., M-1}.

*2) State Preparation:* Alice and Bob prepare coherent states that are initially synchronized in phase and frequency. The mean photon number is independently chosen by Alice and Bob from the set {u, v, w}, where v and w correspond to decoy states and u corresponds to the signal state. The state is then phase-modulated to obtain $|\alpha_A e^{i(a_A+b_A+c_A)}\rangle$ and $|\alpha_B e^{i(a_B+b_B+c_B)}\rangle$, where $a_{A(B)}$ is randomized phase values that determine the phase slice number they belong to, $b_{A(B)}$ corresponds to bases (0 for Z-basis and π/2 for X-basis), and $c_{A(B)}$ represent the phases where key bits are encoded (0 for bit-0 and π for bit-1). Subscripts A and B refer to Alice and Bob states respectively. Z-basis data are used for key generation, and X-basis is utilized for calculating the Quantum Bit Error Rate (QBER).

*3) Transmission:* The prepared states are sent to Charlie, who passes them through a 50:50 beam splitter and publicly announces the detector clicks (D0 and D1) corresponding to each interference.

*4) Iteration:* Steps 2 and 3 are repeated N times for the entire communication block. Subsequently, Alice and Bob announce the mean photon number, bases, and phase slice number (k) used for all N bits communicated.

*5) Post-Selection:* Alice and Bob discard runs with mis-matched phase slices from Charlie's announcement and events where both detectors clicked or none of the detectors clicked. They perform post-selection on the matched events, considering Charlie's announcement for each run to obtain the sifted key. If a click is observed in D0, Alice and Bob assume the same key was transmitted; if D1 clicked, they infer that a different key was transmitted, and Bob flips the bit he had sent in that particular run.

Considering the 50:50 beam splitter of Charlie with phase shift π for the reflected port, the input state on beam splitter is $|\alpha e^{i\gamma_A}\rangle_0 |\alpha e^{i\gamma_B}\rangle_1$, output state is given as:

$$\left|\frac{\alpha e^{i\gamma_A}}{\sqrt{2}} + \frac{\alpha e^{i\gamma_B}}{\sqrt{2}}\right\rangle_{D_0} \left|\frac{\alpha e^{i\gamma_A}}{\sqrt{2}} - \frac{\alpha e^{i\gamma_B}}{\sqrt{2}}\right\rangle_{D_1} \quad (1)$$

with, $\gamma_A = a_A + b_A + c_A$ and $\gamma_B = a_B + b_B + c_B$

In the case of matched runs with same phase slice $\left(a_A \approx a_B, \max(a_A - a_B) = \frac{2\pi}{M}\right)$, same basis ($b_A = b_B$) and if $c_A = c_B$ then $\gamma_A \approx \gamma_B$ in (1). Hence, most of the time detector $D_0$ clicks and few times when $D_1$ clicks, it results in error. The non-zero difference in $(a_A - a_B)$ within the same phase slice causes intrinsic error in the QBER. The contribution of this to the QBER is given as:

$$E_M = \frac{M}{2\pi} \int_0^{\frac{2\pi}{M}} \sin^2\left(\frac{t}{2}\right) dt = 0.5 - \frac{\sin\left(\frac{2\pi}{M}\right)}{\frac{4\pi}{M}} \quad (2)$$

### B. Secret Key Rate

Following the formalism described in [6-8], the lower bound on secret key rate is given by:

$$R \geq \frac{1}{M}\left[-Q_\mu f H(E_\mu) + Q_1(1 - H(e1))\right] \quad (3)$$

where, $Q_\mu$ is the gain, which is the ratio of detected events to the total number of events $\left(\frac{N_{click}}{N}\right)$, $E_\mu$ is the QBER, $f$ accounts for efficiency of error correction, which we consider to be $f=1$ in our analysis, $e1$ is the QBER of detection corresponding to single photon events and H(x) is binary Shannon entropy function. The value $\mu$ is the total mean photon number, equals to sum of mean photon number sent by Alice and Bob ($\mu = \mu_A + \mu_B$) [6].

The value of $Q_\mu$, $E_\mu$ and $e1$ are described as [6-7]:

$$Q_\mu = Y_0 + 1 - e^{-\eta\mu} \quad (4a)$$

$$E_\mu = 0.5 + \frac{0.5}{Q_\mu}(1 - P_{dc})\{e^{-\eta\mu(1-E_m)} - e^{-\eta\mu E_m}\} \quad (4b)$$

$$e_1 = \frac{E_{v_1}Q_{v_1}e^{v_1} - E_{v_2}Q_{v_2}e^{v_2}}{(v_1 - v_2)Y_1} \quad (4c)$$

where $\eta = \eta_{trans} \cdot \eta_{det}$ accounts for product of channel transmittance, $\eta_{trans} = 10^{-\frac{\alpha L}{2 \times 10}}$, with $\alpha$ the channel attenuation coefficient in dB/km and $\eta_{det}$ is the detector efficiency. $Y_0$ and $Y_1$ are yields of no-photon and single photon events respectively, and $E_m$ is the error due misalignments with $v_1$ and $v_2$ are mean photon numbers of decoy state. The factor of 1/M in equation (3) signifies the probability of matching of the phase slice. It should be noted that this equation denotes the asymptotic secret key rate considering that mostly Z-basis and signal state mean photon numbers are transmitted.

The equation for the secret key rate can be expressed as $R \geq [I_{AB} - I_E]$, where $I_{AB}$ is the information shared between Alice and Bob and $I_E$ is the information leaked to Eavesdropper. $I_E$ and $I_{AB}$ in equation (3) are given as: $I_{AB} = \frac{1}{M}\left[Q_\mu\left(1 - fH(E_\mu)\right)\right]$ and $I_E = \frac{1}{M}\left[Q_\mu\left\{1 - \frac{Q_1}{Q_\mu}(1 - H(e1))\right\}\right]$. This considers a pessimistic case in writing out $I_E$ with only the single photon events being untagged and contributes no information to eavesdropper.

### C. Binary Symmetric Erasure Channel Model

The expression of $I_{AB}$ can also be obtained by using binary symmetric erasure channel (BSEC) model. For all the events with matched phase slice, there are three possible outcomes of detector. This includes: (i) either $D_0$ or $D_1$ clicks corresponding to the encoded bits by Alice and Bob, (ii) either both detectors click or none of the detectors click causing erased state or ambiguous detection, meaning no conclusion about sent bit can be made, and (iii) $D_0$ clicks when Alice and Bob encode different bits and vice versa which causes errors in the detection. The maximum information which can be transmitted through the BSEC is:

$$I_{max}^{erasure} = t\log_2 t + (1-t-\epsilon)\log_2(1-t-\epsilon) - (1-\epsilon)\log_2\frac{1-\epsilon}{2} \quad (5)$$

Here, $t$ is the probability of error and $\epsilon$ is the probability of getting ambiguous state. The output state of beam splitter is as given in equation (1). Considering the matched phase slice, matched basis, and same encoded bits, $\gamma_A = \gamma_B + m$, where $m$ indicates the value of phase mismatch $\left(m \in \left[0, \frac{2\pi}{M}\right)\right)$. The mean photon numbers for the output states in (1) are $\mu_{D_0} = 2|\alpha|^2 \cos^2\frac{m}{2}$ and $\mu_{D_1} = 2|\alpha|^2 \sin^2\frac{m}{2}$. The average probability of error is given as:

$$t = \frac{M}{2\pi}\int_0^{\frac{2\pi}{M}} Q_{\mu_{D_1}}\left(1 - Q_{\mu_{D_0}}\right) dm \quad (6a)$$

The average probability of ambiguous result is given as:

$$\epsilon = \frac{M}{2\pi}\int_0^{\frac{2\pi}{M}} \left\{Q_{\mu_{D_0}}Q_{\mu_{D_1}} + \left(1 - Q_{\mu_{D_0}}\right)\left(1 - Q_{\mu_{D_1}}\right)\right\} dm \quad (6b)$$

$Q_\mu$ is the gain as described in [7] which indicates the probability of getting clicks in the detector if coherent state with mean photon number $\mu$ is incident on it. Accounting for the probability of matching of phase slice and solving equation (5) numerically gives: $\frac{1}{M}I_{max}^{erasure} = I_{AB}^{erasure} \approx I_{AB}$.

### III. FOUR DETECTORS IN TF-QKD SETUP

From equation (3), the bound on the secret key rate reveals that two detectors coupled with a single beam splitter, contribute to key generation for the same phase slice selection by Alice and Bob, with a probability of 1/M. Introducing the novel concept of employing four detectors in a specific combination of beam splitters enhances the occurrence of desired events increasing the probability to 4/M. This approach provides a probability distribution of clicks of different detectors, furnishing a distinctive signature to detect the presence of a Beam Splitting (BS) attack during communication.

Figure 2 illustrates the configuration of a 50:50 beam splitter with phase shifts $\phi$ for different beam splitters, where $\phi_{BS1} = \phi_{BS2} = \phi_{BS3} = 0$ and $\phi_{BS4} = \frac{\pi}{2}$. Replacing the detection scheme at Charlie's end from a single beam splitter to a combination of beam splitters and four detectors, the input state in the 4x4 port network is represented as: $|0\rangle_0 |\alpha_A e^{i\gamma_A}\rangle_1 |\alpha_B e^{i\gamma_B}\rangle_2 |0\rangle_3$, and the output state is given by:

$$\left|\frac{\alpha_A e^{i\gamma_A} + \alpha_B e^{i\gamma_B}}{2}\right\rangle_{D_0} \left|\frac{\alpha_A e^{i\gamma_A}(1-i) + \alpha_B e^{i\gamma_B}(1+i)}{2\sqrt{2}}\right\rangle_{D_1}$$

$$\left|-\frac{\alpha_A e^{i\gamma_A}(1-i) + \alpha_B e^{i\gamma_B}(1+i)}{2\sqrt{2}}\right\rangle_{D_2} \left|\frac{\alpha_A e^{i\gamma_A} - \alpha_B e^{i\gamma_B}}{2}\right\rangle_{D_3} \quad (7)$$

If Alice and Bob choose the same mean photon number, then $\alpha_A = \alpha_B = \alpha$, there exists four cases based on difference between $\gamma_A$ and $\gamma_B$. The deterministic events can be interpreted based on clicking pattern of four detectors as shown in Table I. It is noteworthy that for the given difference between $\gamma_A$ and $\gamma_B$, three detectors click while one does not. However, the probability of one detector clicking is higher due to higher mean photon number in one output state. Possible detector clicks based on difference between $\gamma_A$ and $\gamma_B$ for same mean photon number is shown in Table I. In accordance with TF-QKD protocol, where the matching of phase slices is considered, additional phase differences, $m$ may occur apart from the given phase difference in Table I. Therefore, during actual runs, there is a possibility of all four detectors can either show the click or not, resulting in a total of 16 potential outcomes based on their clicking events. These events can be categorized into three different outcomes, namely the desired or correct events, erroneous events, and ambiguous events.

For instance, consider an event where Alice and Bob choose phase slices separated by $\frac{\pi}{2}$, (for M=16, their phase slice number differs by 4, and $a_A \approx a_B + \pi/2$). Moreover, they select the same basis (i.e. $b_A = b_B$) for the same event. The values of $\gamma_A$ and $\gamma_B$ from equation(1a) are given as $\gamma_A \approx a_B + \frac{\pi}{2} + b_B + c_A$ and $\gamma_B \approx a_B + b_B + c_B$. Consequently, $\gamma_A - \gamma_B \approx \frac{\pi}{2} + (c_A - c_B)$ and $(c_A - c_B)$ can be either 0 or $\pi$, which makes $\gamma_A - \gamma_B \approx \pi/2$ or $-\pi/2$ respectively. The interpretation is contingent upon detector clicking, i.e. if $D_2$ clicks but not $D_1$ then as per Table I, $\gamma_A - \gamma_B \approx -\pi/2$. Alice and Bob deduce that their encoded bits are different. Similarly, they interpret the encoded bits based on detector clicks for other phase slice separations with the same chosen basis values, as depicted in Table II. This demonstrates that the number of phase slice matching events have quadrupled. The slight mismatch in random phase components, $a_A$ and $a_B \in \left[0, \frac{2\pi}{M}\right)$ for all the four phase matching slices if $M$ is multiple of four. The analysis for the 4-detector case considered in the paper takes M to be multiple of 4. Due to mismatch in random phase components, detection is also prone to intrinsic QBER as in equation (2). The average value of QBER is given as:

$$E_M^{4\,det} = \frac{M}{2\pi}\int_0^{\frac{2\pi}{M}} \frac{1}{2}\sin^2\left(\frac{t}{2}\right)dt = 0.25 - \frac{\sin\left(\frac{2\pi}{M}\right)}{\frac{8\pi}{M}} \quad (8)$$

TABLE I. OUTPUT STATE BASED ON PHASE DIFFERENCE BETWEEN STATES OF ALICE AND BOB ($\gamma_A - \gamma_B$) AND POSSIBLE DETECTION AT DETECTORS ( ✓- CLICKS, AND ✗- DON'T CLICK )

| $\gamma_A - \gamma_B$ | OUTPUT STATE | $D_0$ | $D_1$ | $D_2$ | $D_3$ |
|---|---|---|---|---|---|
| 0 | $\|\alpha e^{i\gamma_B}\rangle_{D_0} \left\|\frac{\alpha e^{i\gamma_B}}{\sqrt{2}}\right\rangle_{D_1} \left\|\frac{i\alpha e^{i\gamma_B}}{\sqrt{2}}\right\rangle_{D_2} \|0\rangle_{D_3}$ | ✓ | ✓ | ✓ | ✗ |
| $\pi$ | $\|0\rangle_{D_0} \left\|\frac{i\alpha e^{i\gamma_B}}{\sqrt{2}}\right\rangle_{D_1} \left\|\frac{\alpha e^{i\gamma_B}}{\sqrt{2}}\right\rangle_{D_2} \|\alpha e^{i\gamma_B}\rangle_{D_3}$ | ✗ | ✓ | ✓ | ✓ |
| $\frac{\pi}{2}$ | $\left\|\frac{(1+i)\alpha e^{i\gamma_B}}{2}\right\rangle_{D_0} \left\|\frac{(1+i)\alpha e^{i\gamma_B}}{\sqrt{2}}\right\rangle_{D_1} \|0\rangle_{D_2} \left\|\frac{(1-i)\alpha e^{i\gamma_B}}{2}\right\rangle_{D_3}$ | ✓ | ✓ | ✗ | ✓ |
| $-\frac{\pi}{2}$ | $\left\|\frac{(1-i)\alpha e^{i\gamma_B}}{2}\right\rangle_{D_0} \|0\rangle_{D_1} \left\|\frac{(1+i)\alpha e^{i\gamma_B}}{\sqrt{2}}\right\rangle_{D_2} \left\|\frac{(1+i)\alpha e^{i\gamma_B}}{2}\right\rangle_{D_3}$ | ✓ | ✗ | ✓ | ✓ |

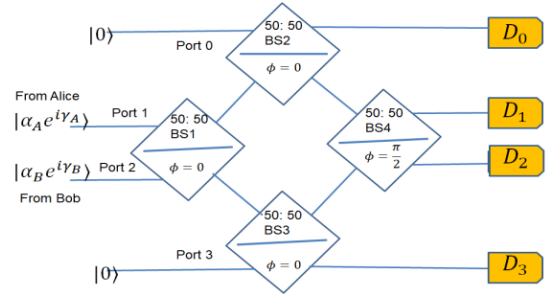

Fig. 2. In the proposed scheme, the detection system at Charlie's end is replaced using beam splitter network of four beam splitters with four detectors with input ports 0 and 3 are set at vaccum state while port 1 and 2 recieves the coherent states coming from Alice and Bob. Detectors are placed in the right named as $D_0$, $D_1$, $D_2$ and $D_3$

TABLE II.  INTERPRETATION OF SIFTED KEY BITS BY ALICE AND BOB DEPENDING ON DETECTOR CLICKS AND ANNOUNCED PHASE SLICE GIVEN BOTH HAVE SELECTED SAME BASIS ( ✓- CLICKS, ✗- DON'T CLICK D – DON'T CARE )

| Random Phase Slice separation | Detector clicks | | | | $c_A$ and $c_B$ same or different ? |
|---|---|---|---|---|---|
| | $D_0$ | $D_1$ | $D_2$ | $D_3$ | |
| $a_A - a_B \approx 0$ | ✓ | D | D | ✗ | Same |
| | ✗ | D | D | ✓ | Different |
| $a_A - a_B \approx \pi$ | ✗ | D | D | ✓ | Same |
| | ✓ | D | D | ✗ | Different |
| $a_A - a_B \approx \pi/2$ | D | ✗ | ✓ | D | Different |
| | D | ✓ | ✗ | D | Same |
| $a_A - a_B \approx -\pi/2$ | D | ✓ | ✗ | D | Different |
| | D | ✗ | ✓ | D | Same |

It is possible that for desired separation of phase slice, desired detector (as shown in Table II) may not register a click, and hence results in an ambiguous detection which is discarded.

A. *Sifted and Secret key rate Extraction*

It is notable that the effective event for key generation, concerning desired phase slice separations as outlined in Table II occurs when a specific detector registers a click. The coherent state output on the port of the desired detector click possesses a mean photon number equivalent to $|\alpha|^2$ (shown in Tables I). In contrast to the scenario with a single beam splitter and two detectors, where the mean photon number on the desired output port is $2|\alpha|^2$ (as per equation (1), if $\gamma_A = \gamma_B$ and $\alpha_A = \alpha_B = \alpha$), the mean photon number on the output port $D_0$ is $\mu = \mu_A + \mu_B = 2|\alpha|^2$. However, with four detectors, the mean photon number on the desired port is $|\alpha|^2$. Therefore, instead of $\mu$, the mean photon number is $\mu/2$, considering $\mu_A = \mu_B = |\alpha|^2$. The information transfer from Alice to Bob, denoted as $I_{AB}^{4\,det}$ is expressed as $\frac{4}{M}\left[Q_{\mu/2}\left(1 - fH(E_{\mu/2})\right)\right]$. Considering $f = 1$, we can derive the information between Alice and Bob using the BSEC model via (5).

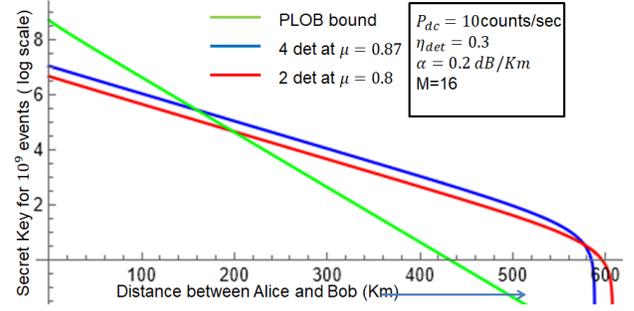

Fig. 3.  Plots of secret key rate vs distance for 2 and 4 detectors setups.

The varied separation in random phase slices (as illustrated in Table II) results in output states that permute among the four output ports. Consequently, the information corresponding to any of the four phase slice separations yields the same value. In this context, we write the state corresponding to the identical slice selection by Alice and Bob including the imperfections in phase due to random phase slice matching, i.e., $a_A - a_B = m$ for $\alpha_A = \alpha_B = \alpha$ as per equation (7) as:

$$\left|\frac{\alpha e^{i\gamma_B}(1+e^{im})}{2}\right\rangle_{D_0} \left|\frac{\alpha e^{i\gamma_B}\{e^{im}(1-i)+(1+i)\}}{2\sqrt{2}}\right\rangle_{D_1}$$
$$\left|\frac{\alpha e^{i\gamma_B}\{-e^{im}(1-i)+(1+i)\}}{2\sqrt{2}}\right\rangle_{D_2} \left|\frac{\alpha e^{i\gamma_B}(-1+e^{im})}{2}\right\rangle_{D_3} \quad (9)$$

The mean photon number at each port is:
$$\mu_{D_0} = \frac{|\alpha|^2}{2}[1+\cos(m)], \mu_{D_1} = \frac{|\alpha|^2}{2}[1+\sin(m)]$$
$$\mu_{D_2} = \frac{|\alpha|^2}{2}[1-\sin(m)], \mu_{D_3} = \frac{|\alpha|^2}{2}[1-\cos(m)] \quad (10)$$

The average probability of error is:
$$t = \frac{M}{2\pi}\int_0^{\frac{2\pi}{M}} Q_{\mu_{D_3}}\left(1 - Q_{\mu_{D_0}}\right)dm \quad (11)$$

The average probability of ambiguous result is:
$$\epsilon = \frac{M}{2\pi}\int_0^{\frac{2\pi}{M}}\left\{Q_{\mu_{D_0}}Q_{\mu_{D_3}} + \left(1-Q_{\mu_{D_1}}\right)\left(1-Q_{\mu_{D_3}}\right)\right\}dm \quad (12)$$

Putting these value in (5), we get $I_{max}^{erasure\,(4\,det)}$ and solving numerically, the expression shows $\frac{4}{M}I_{max}^{erasure\,(4\,det)} = I_{AB}^{erasure(4det)} \approx I_{AB}^{4\,det}$.

Now for the estimation of information leaked $I_E^{4\,det}$, similar argument as used for the scenario of two detectors as in [6] is extended to the four detector case. The valid single photon detection event occurs only if single photon gets detected at the desired detector out of three possible detectors as per Table II.

The probability that single photon gets detected at desired detector = 0.5, and the probability that detector observes click due to single photon =$Y_1\mu e^{-\mu}$ ($Y_1$ is the yield of single photon). Therefore, $I_E^{4\,det} = \frac{4}{M}\left[Q_{\mu/2}\left\{1 - \frac{Y_1\mu e^{-\mu}}{2Q_{\mu/2}}\left(1 - H(e1^{4\,det})\right)\right\}\right]$.

Now secret key rate is given by:
$$R \geq \frac{4}{M}\left[-Q_{\frac{\mu}{2}}fH\left(E_{\frac{\mu}{2}}^{4\,det}\right) + \frac{Y_1\mu e^{-\mu}}{2}(1 - H(e1^{4\,det}))\right] \quad (13)$$

Fig 3. shows the plot for secret key rate after maximizing it for value of $\mu = \mu_A + \mu_B$ at zero length with dark counts taken as $10^{-8}$ corresponding to 1 ns pulse width, $\eta_{det} = 0.3$, channel attenuation coefficient, $\alpha = 0.2$ dB/Km, M=16 and decoy state mean photon numbers are: $v = 0.01$ and $w = 0.001$. Secret key rate for four detector case clearly shows higher value compared to the two detectors. The probability of phase slice selection increases four-fold in the four-detector scenario compared to the two-detector case, and consequently the secret key rate increases by only 2.3 times. This discrepancy arises from higher probability of ambiguous events in the four-detector setup. It is worth noting that two slice selections are also possible in the two-detector case, though not implemented in the original proposal [6]. Including the additional phase slice doubles the key rate. Consequently, the advantage in key rate increment for four detectors is expected to be 1.15 times higher than that for the two detector case with two phase slice selections. The advantage an eavesdropper can gain from exploiting the announcement of the global phase for a collective beam splitting attack is not considered into the secret key rate. As detailed in [6], the maximum information that can be extracted is upper bounded by the Holevo bound, which remains the same for both two-detector and four-detector cases. Even after reducing this extra information leakage, the system still demonstrates better performance than the PLOB bound.

*B. Determining BS Attacks using Detector click Distribution*

The four-detector setup offers an additional advantage in detecting the signature of BS attacks. It is worth noting that if an eavesdropper attempts to intercept photons from both Alice and Bob to Charlie, it's crucial to ensure an equal fraction of photons are intercepted from both sides. Otherwise, this imbalance can lead to an increase in the QBER. However, with careful adjustment to ensure equal fraction of photons are intercepted from both sides without disturbing the transmitted pulse's phase, the BS attack won't affect the QBER. Generally, BS attack assumes the use of lossless channel by eavesdropper which enables him to use beam splitter having transmittance equal to channel transmittance. This kind of attack scenario is difficult to detect as Alice and Bob assumes the loss is expected due to channel. The incorporation of four detectors does not help in detecting such type of attack. However, if the eavesdropper cannot achieve a lossless channel and attempts to split the photon beam from both arms using identical beam splitters, this can be detected with the four-detector setup.

With the four numbers of detectors, there are five potential outcomes for each event of sent photon pulses, namely: (i) no detectors click, (ii) one detector clicks, (iii) two detectors click, (iv) three detectors click, and (v) all detectors click. These events of detector click follow a distribution influenced by the channel, detector efficiency, dark count, mean photon number of the photon pulse, and the fraction of photons intercepted by the eavesdropper $(r)$. With a known communication channel, parameters of detector and mean photon number, Alice and Bob anticipate a specific distribution of detector click events, which is disrupted by a BS attack. If the eavesdropper manipulates Charlie and falsely reports detector's click in addition to the actual observed clicks to offset the expected detector's click, it leads to an increase in QBER. This is because one of the four detectors must not click, and the announcement of detector's click precedes the announcement of phase slice. Moreover, the probability of no detector click increases with an increase in $r$. If Charlie feigns no click in a detector, the event will be discarded by Alice and Bob.

The distribution with respect to intercepted fraction is derived using the mean photon number in equation (10), replacing $|\alpha|^2$ by $(1-r)^2|\alpha|^2$, and hence changing the notation of $\mu_{D_i}$ to $\mu_{D_i}^r$. While the distribution can be obtained for all events even if the phase slices do not match as per Table II, it is useful to consider events from matched phase slices only to detect if Charlie falsely announces the click of extra detectors.

Average Probability of no detectors click, $P_{NC}$:
$$P_{NC} = \frac{M}{2\pi} \int_0^{\frac{2\pi}{M}} (1-Q_{\mu_{D_0}^r})(1-Q_{\mu_{D_1}^r})(1-Q_{\mu_{D_2}^r})(1-Q_{\mu_{D_3}^r}) \, dm \quad (14)$$

Average Probability of one detector clicks, $P_{OC}$:
$$P_{OC} = \frac{M}{2\pi} \int_0^{\frac{2\pi}{M}} \{Q_{\mu_{D_0}^r}(1-Q_{\mu_{D_1}^r})(1-Q_{\mu_{D_2}^r})(1-Q_{\mu_{D_3}^r}) + (1-Q_{\mu_{D_0}^r})Q_{\mu_{D_1}^r}(1-Q_{\mu_{D_2}^r})(1-Q_{\mu_{D_3}^r}) + (1-Q_{\mu_{D_0}^r})(1-Q_{\mu_{D_1}^r})Q_{\mu_{D_2}^r}(1-Q_{\mu_{D_3}^r}) + (1-Q_{\mu_{D_0}^r})(1-Q_{\mu_{D_1}^r})(1-Q_{\mu_{D_2}^r})Q_{\mu_{D_3}^r}\} \, dm \quad (15)$$

Average Probability of two detectors click, $P_{TC}$:
$$P_{TC} = \frac{M}{2\pi} \int_0^{\frac{2\pi}{M}} \{Q_{\mu_{D_0}^r}Q_{\mu_{D_1}^r}(1-Q_{\mu_{D_2}^r})(1-Q_{\mu_{D_3}^r}) + Q_{\mu_{D_0}^r}(1-Q_{\mu_{D_1}^r})Q_{\mu_{D_2}^r}(1-Q_{\mu_{D_3}^r}) + Q_{\mu_{D_0}^r}(1-Q_{\mu_{D_1}^r})(1-Q_{\mu_{D_2}^r})Q_{\mu_{D_3}^r} + (1-Q_{\mu_{D_0}^r})Q_{\mu_{D_1}^r}Q_{\mu_{D_2}^r}(1-Q_{\mu_{D_3}^r}) + (1-Q_{\mu_{D_0}^r})Q_{\mu_{D_1}^r}(1-Q_{\mu_{D_2}^r})Q_{\mu_{D_3}^r} + (1-Q_{\mu_{D_0}^r})(1-Q_{\mu_{D_1}^r})Q_{\mu_{D_2}^r}Q_{\mu_{D_3}^r}\} \, dm \quad (16)$$

Average Probability of three detectors click $P_{THC}$:
$$P_{THC} = \frac{M}{2\pi} \int_0^{\frac{2\pi}{M}} \{Q_{\mu_{D_0}^r}Q_{\mu_{D_1}^r}Q_{\mu_{D_2}^r}(1-Q_{\mu_{D_3}^r}) + Q_{\mu_{D_0}^r}Q_{\mu_{D_1}^r}(1-Q_{\mu_{D_2}^r})Q_{\mu_{D_3}^r} + Q_{\mu_{D_0}^r}(1-Q_{\mu_{D_1}^r})Q_{\mu_{D_2}^r}Q_{\mu_{D_3}^r}\} \, dm \quad (17)$$

Average Probability of four detectors click $P_{FC}$:
$$P_{NC} = \frac{M}{2\pi} \int_0^{\frac{2\pi}{M}} Q_{\mu_{D_0}^r}Q_{\mu_{D_1}^r}Q_{\mu_{D_2}^r}Q_{\mu_{D_3}^r} \, dm \quad (18)$$

To assess the similarity between the observed distribution and the expected distribution, $\chi^2$-test can be employed to validate the presence of BS attack.

## IV. COMPARISON WITH SIMULATIONS

The validity of theoretical modeling is confirmed through simulations using python-based Strawberry-Fields package [10], for state preparation, circuit definition, and fock state measurement. Two coherent states are prepared with identical phase and mean photon numbers. Sixteen equal phase slices are then prepared, and uniformly distributed random numbers ranging from 0 to $2\pi$ are generated to set the values of $a_A$ and $a_B$. The phase slice numbers for both random numbers are recorded. Additionally, two binary random numbers are generated to set the bit values ($c_A$ and $c_B$). For simplicity, the simulation assumes that both Alice and Bob are using the same basis values, so separate random values for deciding the basis ($b_A$ and $b_B$) are not considered. Subsequently, the prepared coherent states are sent to beam splitters, and their output is measured. To account for the detector efficiency and channel transmittance, a uniform random number is generated between 0 and 1. If the random number is greater than $1-(1-\eta)^n$, the variable corresponding to detector click is set to zero, indicating no detection occurred; otherwise, it is set to one. $n$ is the number of photons measured.

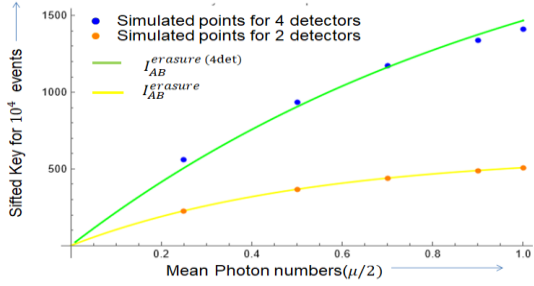

Fig. 4. Graph comparing the sifted key rate with respect to mean photon number between two and four detectors with theoretical model and simulated data for $10^4$ events for ideal detectors and no channel loss.

A search for the desired events of matched phase slices is conducted, and the event is considered for key generation if the detection pattern meets the specified requirements. A simulation involving $10^4$ events was conducted under ideal channel and detector conditions considering a total of 16 phase slices for both the two-detector and four-detector scenarios. The variation of the sifted key rate with respect to changes in mean photon numbers, $\mu_A = \mu_B = \mu/2$ is depicted in Fig. 4. Notably, the simulated model aligns closely with the theoretical results confirming that the achieved sifted key rate for the four-detector setup surpasses that of the two-detector configuration. Furthermore, another simulation carried out for both the two-detector and four-detector setups, comprising $10^5$ events, while varying the distance between Alice and Bob. Charlie is assumed to be positioned at the midpoint between them. The parameters set for this simulation include a mean photon number, $\mu = \mu_A + \mu_B$ of 0.5, attenuation coefficient of 0.2 dB/km for the channel, $\eta_{\text{det}} = 1$ with zero dark count. The simulation results are illustrated in Fig. 5. Each point on the graph represents the outcome of the simulation, while the solid line that accompanies the points represents $I_{AB}^{erasure}$ and $I_{AB}^{erasure}$ (4det). The curve clearly indicate that the key rate increases with the adoption of four detectors.

## V. DISCUSSIONS

The proposed idea of using four detectors shows promising results in increasing the key rate and detecting beam splitting

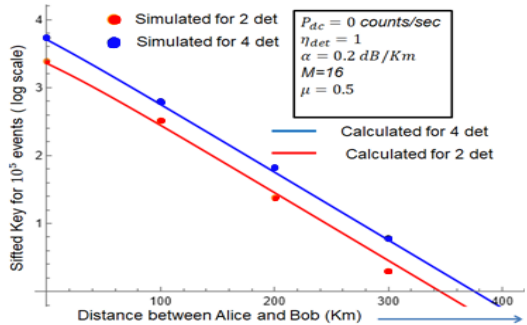

Fig. 5. Graph of sifted key for $10^5$ events for two and four detectors simulated (Dots) and theoretical calculation (Solid lines) at various distances

(BS) attacks. One of the main challenges in the practical implementation of TF-QKD is the synchronization of the coherent sources of Alice and Bob, which has been addressed by various methods [11-12]. The involvement of four detectors, however, increases the cost and complexity of the system. The calculations used in this paper assume the same parameters for all detectors and that the beam splitters used in the network are lossless. In the case of lossy beam splitters, the parameters responsible for loss need to be considered in estimation of $\eta$ in equation (4). Detectors D0 and D3 experience photon detections after passing through two beam splitters, while detectors D1 and D2 experience photon detections after passing through three beam splitters. Therefore, a separate analysis corresponding to events with detections at different detectors needs to be performed if non-idealities in the beam splitters and detectors with different parameters are considered.

## VI. CONCLUSION

The use of four detectors in conjunction with a specific combination of four beam splitters enhances the key rate by increasing the probability of selecting the desired phase slice, albeit with a slight decrease in achievable distance. Additionally, the probability distributions of detector click with four detectors provide an effective method for detecting beam splitting (BS) attacks.